\documentclass[lettersize,journal]{IEEEtran}
\ifCLASSINFOpdf
\else
\fi

\usepackage{graphicx} % Required for inserting images
\usepackage{amsmath}
\usepackage{xcolor}
\usepackage{algorithm} 
\usepackage{algpseudocode}
\usepackage{setspace}
\usepackage[most]{tcolorbox}
\usepackage{xcolor}
\usepackage{multirow}
\usepackage{longtable}
\usepackage{tabularx}
\usepackage{array}
\usepackage{hyperref}

\usepackage{longtable}
\usepackage{array}
\usepackage{multirow}

\usepackage{listings}
\usepackage{xcolor}

\definecolor{codebg}{RGB}{245,245,245}
\definecolor{keyword}{RGB}{33,74,135}
\definecolor{comment}{RGB}{120,120,120}
\definecolor{string}{RGB}{160,60,60}

\lstdefinelanguage{Solidity}{
  keywords={contract, address, mapping, public, private, uint, constructor, function, for, require},
  keywordstyle=\color{keyword}\bfseries,
  ndkeywords={send, push},
  ndkeywordstyle=\color{keyword},
  identifierstyle=\color{black},
  sensitive=true,
  comment=[l]{//},
  commentstyle=\color{comment}\itshape,
  stringstyle=\color{string},
  morestring=[b]"
}

\title{Evaluating the Vulnerability Landscape of LLM-Generated Smart Contracts}

\begin{document}

\newcolumntype{L}[1]{>{\raggedright\arraybackslash}p{#1}}

\newcommand{\nasrin}[1]{\textcolor{red}{#1}}
\newcommand{\muneeb}[1]{\textcolor{magenta}{#1}}
\newcommand{\TODO}[1]{\textcolor{blue}{#1}}
\newcommand{\hoang}[1]{\textcolor{brown}{#1}}
 \author{
  {\rm Hoang Long Do$^\ast$, Nasrin Sohrabi$^\ast$, Muneeb Ul Hassan$^\ast$}\\
  $^\ast$ Deakin University, Australia
  }

\maketitle

\begin{abstract}

Large language models (LLMs) have been widely adopted in modern software development lifecycles, where they are increasingly used to automate and assist code generation, significantly improving developer productivity and reducing development time. In the blockchain domain, developers increasingly rely on LLMs to generate and maintain smart contracts, the immutable, self-executing components of decentralized applications. Because deployed smart contracts cannot be modified, correctness and security are paramount, particularly in high-stakes domains such as finance and governance. Despite this growing reliance, the security implications of LLM-generated smart contracts remain insufficiently understood.

In this work, we conduct a systematic security analysis of Solidity smart contracts generated by state-of-the-art LLMs, including ChatGPT, Gemini, and Sonnet. We evaluate these contracts against a broad set of known smart contract vulnerabilities to assess their suitability for direct deployment in production environments. Our extensive experimental study shows that, despite their syntactic correctness and functional completeness, LLM-generated smart contracts frequently exhibit severe security flaws that could be exploited in real-world settings. We further analyze and categorize these vulnerabilities, identifying recurring weakness patterns across different models. Finally, we discuss practical countermeasures and development guidelines to help mitigate these risks, offering actionable insights for both developers and researchers. Our findings aim to support safe integration of LLMs into smart contract development workflows and to strengthen the overall security of the blockchain ecosystem against future security failures.

\end{abstract}

\section{Introduction}

Large language models (LLMs) have rapidly become integral to modern software development lifecycle, where they are increasingly used to automate and assist code generation, debugging, and maintenance, substantially improving developer productivity and reducing development time \cite{mhint09}. This paradigm shift is now extending to the blockchain domain, where LLMs are being actively leveraged to generate smart contracts, the core building blocks of decentralized applications (DApps). By lowering the technical barrier to entry, LLMs enable even non-expert developers to produce functionally complete smart contracts, accelerating development and fostering broader adoption of blockchain-based systems. However, this growing reliance on LLM-generated code raises fundamental questions regarding the security and reliability of the resulting smart contracts, particularly given their immutable and autonomous execution semantics.

Modern blockchain platforms are built around smart contracts, which serve as a fundamental abstraction for implementing self-executing programs that autonomously enforce predefined conditions on decentralized infrastructures \cite{mhint03, mhint04}. The modern notion of smart contracts was first realized/introduced by Ethereum with the introduction of the Ethereum Virtual Machine (EVM) in 2015, which allowed the execution of programmable logic in a decentralized and autonomous manner \cite{mhint05}. This innovation facilitated the emergence of DApps, in which contract logic is enforced directly by blockchain rather than by trusted intermediaries. Smart contracts allow developers to embed executable logic into blockchain ledger, supporting a wide range of advanced functionalities, including: digital asset custody, access control enforcement, auction and market mechanisms, and decentralized governance coordination, all without reliance on centralized authorities \cite{mhint08}. Typically written in high-level languages such as Solidity and deployed on decentralized blockchain networks, smart contracts provide transparency and auditability while significantly reducing the need for manual policy enforcement.

However, the immutability and autonomous execution of smart contracts also highlights the impact of software vulnerabilities, as any flaw in the contract logic may be irrevocably exploited once deployed, underscoring the critical importance of security analysis prior to deployment. A well-known example is the 2016 DAO hack, which resulted in the loss of millions of dollars and highlighted the severe consequences of insecure smart contract design~\cite{mhint07}. To mitigate such catastrophic failures, rigorous security practices, such as formal verification, systematic auditing, and adherence to widely adopted standards (e.g., ERC-20 for tokens), have become essential prior to deployment. These measures aim to detect and prevent common classes of vulnerabilities and errors, including reentrancy, overflow errors, and access control violations.

The integration of LLMs into smart contract development further complicates this security landscape. While LLMs can rapidly generate syntactically correct and functionally plausible smart contracts, they lack a fundamental understanding of adversarial behaviors, execution semantics, and blockchain-specific threat models. As a result, LLM-generated smart contracts may unintentionally introduce subtle yet critical vulnerabilities that are difficult to detect through superficial inspection. Given the irreversible nature of smart contract deployment, this raises a pressing question: \textit{are LLM-generated smart contracts sufficiently secure for direct deployment in production environments?}

To address this question, we conduct a systematic security analysis of smart contracts generated by state-of-the-art LLMs, including ChatGPT, Gemini, and Sonnet. We generate and analyze smart contracts across multiple application domains, such as decentralized finance, token management, governance and voting, social networks, and marketplaces. Our experimental results demonstrate that LLM-generated smart contracts frequently exhibit serious security vulnerabilities and are unsuitable for direct deployment without extensive auditing and remediation. These findings highlight the urgent need for a deeper understanding of LLM-induced risks in smart contract development and for robust safeguards to ensure secure adoption of AI-assisted programming tools in blockchain ecosystems.

\noindent\textbf{Key Contributions.}
The key contributions of this work are as follows:

\begin{itemize}
    \item We conduct a comprehensive security evaluation of smart contracts generated by leading LLMs, including ChatGPT, Gemini, and Sonnet, demonstrating that such contracts are generally unsuitable for direct deployment without rigorous security review.

    \item We identify and categorize vulnerability patterns specific to LLM-generated smart contracts, providing a foundational taxonomy for understanding LLM-induced risks in blockchain development.
    
    \item We propose a detailed threat model and present an extensive experimental analysis of vulnerability prevalence, generation pipelines, and vulnerability classes across different application domains.
    
    \item  We provide practical insights and mitigation strategies to support the safe integration of LLMs into smart contract engineering workflows.
\end{itemize}

The remainder is paper is organized as follows: Section \S\ref{sec:litreveiw} provides an in-depth literature review. In Section \S\ref{sec:prelim}, we explore the preliminaries of our proposed model. Next in section \S\ref{sec:threatModel}, we discuss the threat model. Afterwards, in section \S\ref{sec:SystMdl}, we demonstrate the system model and function of our model. Then in section \S\ref{sec:ExprEval}, we provide in-depth experimental evaluation. Finally, section \S\ref{conclusion} concludes the article.

\section{Literature Review} \label{sec:litreveiw}

Blockchain technology has matured into foundation infrastructure for decentralized computation, asset management, and trustless coordination because of the advancements in smart contracts and DApps. Multiple studies have demonstrated the effectiveness of smart contracts across real-world domains, including finance, supply chain management, and healthcare. Additionally, recent research highlights a growing trend toward integrating large language models (LLMs) into blockchain development lifecycle to automate and assist smart contract development. However, prior works shows that LLMs are not a one-stop solution for software development~\cite{mhlit07,mhlit08}. As LLM-generated code, particularly in languages such as C++ and Java, frequently contains security vulnerabilities that must be carefully addressed before deployment.

These vulnerabilities become more critical when considered in smart contract scenarios, as these contracts are irreversible and unstoppable once executed. Thus, in this section, we study existing literature from the perspective of advancement in blockchain and LLMs for blockchain and smart contract development alongside highlighting the need that no existing work focused over the vulnerabilities of LLM generated smart contracts. \\

Over the past years, numerous studies have examined the design, security, and practical integration of smart contracts across modern application domains.
A very comprehensive study focusing on highlighting a wide range of security vulnerabilities in smart contracts was presented by Zhu~\textit{et al.}~\cite{mhlit01}. In this work, the authors synthesize evidence from 500+ real-world Ethereum security incidents and demonstrate that smart contracts exhibit multiple classes of security vulnerabilities, which can lead to potential exploits, security breaches, and financial losses. The authors also present various mitigation techniques alongside a range of open challenges and future research directions, such as the need for better benchmarks, stronger cross-contract analysis, and robustness against adversarial behavior.

Another work that proposes a static analysis framework for detection of reentrancy vulnerability in Solidity smart contracts have been presented by authors in~\cite{mhlit02}. The authors highlighted different measures to accurately spot the vulnerability, e.g., where an attacker can ``hook" control flow (entry points), alongside from where the attacker can profit from these attacks (reentry points). The key contribution is towards tracking the flow of delayed-updated state variables with static taint analysis to confirm whether those stale states can actually influence exploitable operations, which helps catch cross-function reentrancy and reduce false positives compared to prior tools. Similarly, another model named VULSEYE was proposed by Liang~\textit{et al.} in~\cite{mhlit03}. This work proposed a stateful directed graybox fuzzer for smart contracts, which first finds vulnerability-prone code targets via static analysis and state targets via backward bytecode analysis. It then guides fuzzing using a fitness score based on distance to code and state targets, improving bug-finding over prior coverage-guided fuzzers.\\

From the perspective of integrating LLMs into programming-related scenarios, Fu~\textit{et al.}~\cite{mhlit04} present a comprehensive literature review on the use of LLMs in programming education. The study provides several valuable insights by proposing a six-dimensional framework, demonstrating that LLMs are evolving from mere programming tools into instructional agents that primarily support resource generation, task solving, and feedback delivery. Moving toward the application of LLMs in blockchain smart contract development, Wei~\textit{et al.}~\cite{mhlit05} propose a framework called LLM-SmartAudit, which focuses on smart contract vulnerability detection through LLM-powered multi-agent collaboration. In their proposed approach, the system assigns specialized agents to collaboratively analyze smart contracts using both broad and targeted analysis modes, enabling more accurate detection of common vulnerabilities as well as complex logic flaws. Furthermore, another related work~\cite{mhlit06} focuses on identifying novel numerical defect classes from smart contract audit reports, targeting vulnerabilities beyond classic overflow-style bugs. This work combines symbolic execution over both bytecode and source-level features with LLM-based function pruning to mitigate path explosion and significantly accelerate vulnerability analysis.

After carefully analyzing the recent literature, to the best of our knowledge, no work focuses on studying the security vulnerabilities produced as a result of LLM-generated smart contracts.

\section{Preliminaries} \label{sec:prelim}
In this section, we focus over the foundational concepts of our proposed mechanism. We first discuss the core principles of blockchain, then we highlight the generation smart contracts via LLM models. Afterwards, we discuss the types of vulnerabilities and then we discuss the adversarial effect of these vulnerabilities alongside the impacts that can be caused if these vulnerabilities go unnoticed. 

\subsection{Blockchain and Smart Contracts}

\subsubsection{Blockchain and its Functioning}
Blockchain technology first came to discussion with the publication of the white paper of Bitcoin in 2008 by Satoshi Nakamoto~\cite{mhref01}. This paper introduced a new peer-to-peer (P2P) decentralized electronic cash system, which was built on a tamper-proof and append-only ledger, maintained by a group of nodes which operate in a decentralized fashion. Bitcoin worked over the phenomenon of maintaining its decentralized ledger via proof-of-work (PoW) consensus, which basically allows decentralized participants to reach a consensus without any centralized mediation~\cite{mhref02}. Initially Bitcoin only attracted the attention of cryptography enthusiast, but due to its unique consensus and trust model, between 2009-2013, it evolved from a niche experiment to a global decentralized currency, which attracted the attention of both academia and industry, and researchers and developers started exploring the notion behind Bitcoin (which was blockchain technology).\\
Around 2013-2015, the focus on blockchain technology shifted from a single-purpose payment models towards a bit more generic computation model. In late 2013, with the advent of decentralized applications (DApps) by Ethereum, a new era of Blockchain also known as Blockchain 2.0 started~\cite{mhref03}. These DApps worked over the notion of smart contracts, which works as self-executing digital agreements that are coded on blockchain technology. These contracts allowed blockchain practitioners to perform various advanced digital operations on blockchain which were not possible on Bitcoin. From 2017 onward, these smart contracts and public blockchain became mature enough towards handling large infrastructure-oriented applications, paving the way for applicability of blockchain in various industrial sectors such as financial markets, governance systems, marketplaces, etc. 

\subsubsection{Smart Contracts}
Smart contracts can be termed as self-executing programs that are deployed on the blockchain technology, with an aim to automatically execute and enforce the terms of agreement in case when the predefined conditions are met. As discussed earlier, smart contracts came into discussion with the advent of DApps by Ethereum in 2013. Since, smart contracts operate on a Turing-compete virtual machine (e.g., Ethereum Virtual Machine (EVM)), thus, they allow complex logic to be encoded in their language (such as Solidity, etc) ~\cite{mhref04}. Smart contracts do not reply on a centralized entity (e.g., bank or an exchange), thus, in case of a development of a smart contract, the corresponding parties encode the desired behaviour (such as making new registration, transferring tokens, etc) into the contract and process it further for the deployment once required.\\ 
The typical life cycle of smart contract is pretty simple, as the developer writes the smart contract in a high-level language (e.g., Solidity), then the developer compiles this contract into bytecode, and then deploy it to the chain by sending a special transaction, which basically creates a new contract on the blockchain architecture. From this point onward, the users or other contracts can send transactions to trigger the functions/requirements of smart contract in order to execute it. Since, deployed smart contracts are nearly impossible to change, thus, even a small mistake or vulnerability in their logic can have severe consequences on the blockchain environment~\cite{mhref05}. This basically builds the foundation of our work that if the developer generates some smart contract via LLMs and deploy it directly without checking for vulnerabilities, then it can lead to catastrophic outcome (more about this in the upcoming sections).

\subsection{AI-Generated Smart Contracts}
Since the emergence of ChatGPT and other LLMs in 2021, they have become a mainstream tool for majority of development related tasks. Github introducing ``Github Copilot" in mid-2021 started a new era of pairing AI with programming, which started from basic code completion and has now taken over a major chunk of developing work. In practice, a large number of developers regularly use LLMs to write new programs, translate between programming languages, write the tests, developing quick prototypes, etc. Similar to this, smart contract development also followed the same trend, and many recent works proposed and used various LLMs to generate smart contracts for multiple application and development environment, such as Ethereum, Hyperledger Fabric, etc~\cite{mhref06}. 

\subsubsection{Generation Pipeline}
The generation pipeline for LLM based smart contracts mainly involve three key steps. Firstly, the developer write a high-level description of the required contract in natural language prompt. Prompts can be as simple as plain text or can include code fragments, key constraints, etc. After this, the prompt is sent to an LLM (LLM can be a general-purpose or a code-specialized model) through an API call or via chat interface. The LLM model typically generates the smart contract by following the instructions given in the prompt. Lastly, the generated code is then checked for tests, compiled, and then deployed on the blockchain architecture. In practice, many developers do stop after a single generation if the code ticks all the basic functional tests, which basically raises various security concerns. 

\subsubsection{LLM Models Under Study}
In our experiments, we employ three general-purpose LLMs, GPT-4.1, Gemini-2.5, and Claude Sonnet-4.5, which are trained on mixed natural-language and code corpora, rather than models specialized for blockchain development. We deliberately select general-purpose LLMs because they are more commonly used by everyday developers in practice, in contrast to domain-specific models that see more limited adoption. In this study, we treat all selected models as black-box code generators: each model is provided with a prompt, produces corresponding source code, and the generated programs are subsequently analyzed for security vulnerabilities.

\subsubsection{Characteristics of AI-Generated Smart Contract}
Smart contracts generated using AI exhibit several distinct characteristics that differentiate them from human-written smart contracts. From a positive perspective, AI-generated smart contracts often adhere closely to established standards, such as ERC-20 and ERC-721. In addition, the generated code is typically well structured, readable, and frequently reflects idiomatic patterns commonly observed in open-source repositories and programming tutorials. Moreover, AI-generated smart contracts tend to achieve a high compilation success rate on initial attempts when compared to human-written code.

However, despite these advantages, closer inspection reveals numerous weaknesses in AI-generated smart contracts. In some cases, AI models introduce unnecessary or redundant code fragments that are not required for correct functionality. Furthermore, because LLMs are susceptible to hallucinations, they may occasionally invent non-existent functions or incorrectly apply language primitives and APIs~\cite{mhref07}. In addition, the generated logic is sometimes insufficiently optimized, leading to inefficient gas consumption. Most critically, AI-generated smart contracts may contain severe security vulnerabilities which, if left unaddressed, can result in serious financial and operational consequences.

\subsection{Types of Smart Contract Vulnerabilities}
Smart contracts can fail in multiple ways if not handled correctly, ranging from some low-level programming errors to the subtle violations of protocol-level issues. In here, we categorize these vulnerabilities into two subtypes as discussed further. 

\subsubsection{Classical Smart Contract Vulnerabilities}
Classical vulnerabilities are the longstanding security flaws in smart contracts that have been extensively studied and documented since the advent of smart contracts. These flaws usually stem from the misunderstandings of EVM or solidity semantics~\cite{mhref08}. Some of the most common examples of classical vulnerabilities include reentrancy, arithmetic overflow and underflow, denial of service via gas-execution, and improper initialization of critical variables. Since, these vulnerabilities are well-studied, therefore, they can be detected via formal verifications. Later in this research paper we highlight that AI generated smart contracts produce large number of these vulnerabilities, therefore, it is important to detect and correct them thoroughly.

\subsubsection{Logic Vulnerabilities}

Logic vulnerabilities arise when the implemented code deviates from the intended protocol behavior, even though the code itself is syntactically correct~\cite{mhref10}. Such vulnerabilities are often subtle and highly context-dependent, making their detection particularly challenging. In essence, these flaws can be summarized as cases where the `` program behaves exactly as written", yet the underlying design or logic of the program is incorrect. Common examples include unsafe pricing formulas, flawed fee mechanisms in automated markets, and incorrect liquidation rules. If left uncorrected, logic vulnerabilities can lead to catastrophic financial and operational consequences. Since LLMs are prone to hallucinations, they may introduce such fundamental logical errors during code generation. Identifying and correcting these vulnerabilities therefore requires careful inspection and a deep understanding of the intended protocol semantics.

\subsection{Adversarial Effect of Vulnerabilities in Production}

From an adversarial viewpoint, the vulnerabilities discussed above describe different layers of blockchain smart contract that can be exploited by attackers. Once an AI-generated vulnerable smart contract enters the production stage, it becomes a part of a global and fully observable execution environment. Now, from there onward, any external account can interact with it, and the complete code of smart contract is visible to potential attackers, who can further offline-simulate and stress-test the code in order to achieve their malicious aims. In this part, we examine the capabilities of an adversary along with outlining the impact it can cause.

\subsubsection{Adversary Capabilities}

Adversaries that target the blockchain smart contracts possess a wide range of capabilities, ranging from automated scanners to sophisticated on-chain manipulation tools. Basic attacker may use some static analyzing tool, such as Slither or MythX to detect the classical vulnerabilities. While the advanced attackers can leverage dynamic analysis and fuzzing techniques to identify more complex vulnerabilities. More critically, the advent and integration of AI and LLMs in blockchain technology arose as a double edged sword, on one hand it has helped developers to develop the contracts faster, however on the other hand, it is also helping the malicious entities to identify the vulnerabilities from their end. Any malicious entity can get the code from public blockchain and simply ask the AI model to identify the vulnerabilities in it, and just like it did in the generation phase, the LLM model can also identify the vulnerability in the code. This information can then maliciously be used by adversaries to fulfill their malicious objectives. However, it is also to mention that even if malicious entities attempt an LLM `generate-audit-fix' loop, the LLMs are not a fully dependable security analyzers and can both miss vulnerabilities and hallucinate safety. Hence, LLM based self review from the developers side cannot replace rigorous verification and independent auditing, and does not negate the attacker advantage from public code visibility.

\subsubsection{Impact on Assets and Protocols}
Exploiting a vulnerability on blockchain can lead to serious consequences on digital assets and the whole ecosystem, ranging from financial losses to severe system denial and disruptions. One of the most common immediate effect is the unauthorized movement of assets, such as draining liquidity pools, minting, unlocking tokens, permanently locking user funds. Additionally, utility and registry contracts that are vulnerable to access control breaches can result in the compromising of data integrity, which can also lead to falsified records and unauthorized transfer of assets. From a broader viewpoint, the implications could include network congestion, denial of service, and reputation damage as well. 

Considering, all these factors, it is important to identify that whether the AI generated smart contracts are secure or vulnerable? and if they are vulnerable, what are the most common vulnerabilities among them that needs to be tackled before sending them to production. In the upcoming sections, we discuss all these questions in detail.

\section{Threat Model} \label{sec:threatModel}

The adoption of LLM-based code assistants by inexperienced smart contract developers is rapidly increasing, as these tools significantly accelerate the development process. However, this speed often comes at the cost of security. Even code generated by the most advanced models may contain serious vulnerabilities, and existing auditing practices may be insufficient to detect all flaws prior to deployment.

To capture this realistic setting, we consider an adversarial framework that models the interaction between a programmer and commercial LLM-based coding assistants, such as GitHub Copilot. The programmer provides a high-level specification or intended functionality of a smart contract, which is submitted to the LLM through a chat-based interface. The resulting code is assumed to be reviewed only for \textit{syntactic correctness} before deployment.

\begin{figure}
    \centering
    \includegraphics[width=1\linewidth]{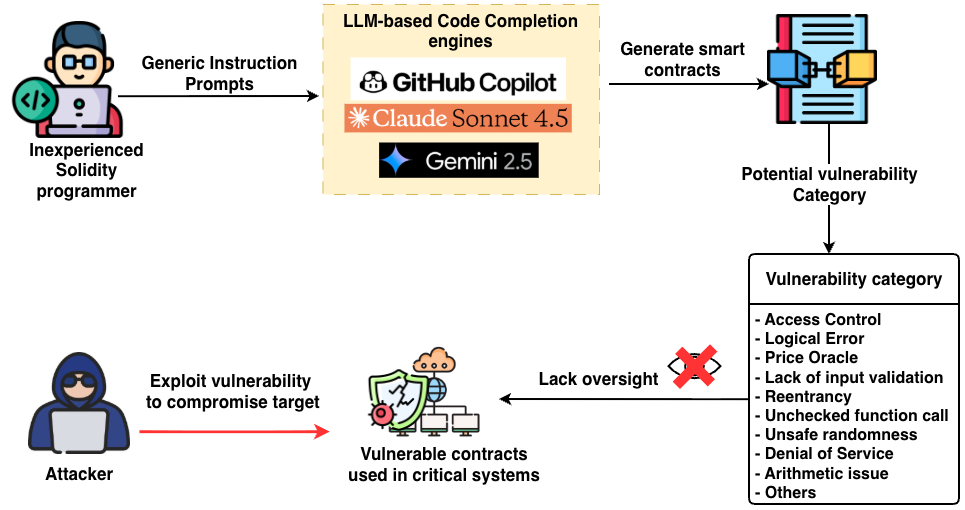}
    \caption{Scenario where security issues caused serious damage to critical systems.}
    \label{fig:threat-model}
\end{figure}

Crucially, syntactic correctness does not imply security; unverified code may contain exploitable vulnerabilities. Adversaries may actively scan deployed contracts for such weaknesses and exploit them to cause significant financial losses. As illustrated in Fig.~\ref{fig:threat-model}, our experimental setup demonstrates how over-reliance on LLM-based code generation can introduce systemic vulnerabilities and expand the attack surface of smart contract ecosystems.

\section{System Model and Functioning} \label{sec:SystMdl}

\subsection{System Overview}

\begin{figure*}
    \centering
    \includegraphics[width=1\linewidth]{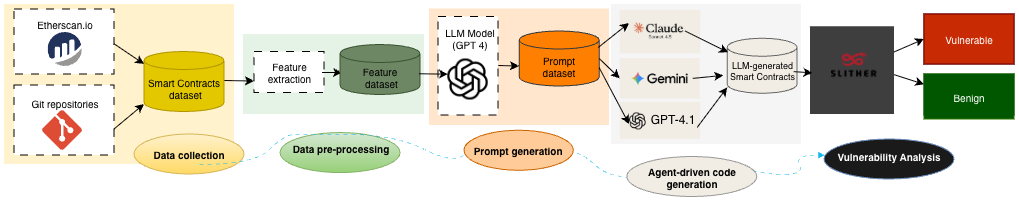}
    \caption{Proposed streamlined system for vulnerability testing of agent-generated smart contracts. It comprises five modules: smart contract collection, feature extraction, prompt-generation, agent-driven code generation, and vulnerability analysis.}
    \label{fig:sysmodel}
\end{figure*}

We propose a streamlined system for evaluating the \textit{reliability} and \textit{correctness} of code generated by LLM-based coding agents, as illustrated in Fig.~\ref{fig:sysmodel}. The system comprises five core modules: \textit{(i)} smart contract collection, \textit{(ii)} feature extraction, \textit{(iii)} prompt-generation dataset construction, \textit{(iv)} agent-driven code generation, and \textit{(v)} vulnerability analysis of the generated code. We provide an overview of each module below. \\

The system begins by collecting smart contracts from publicly accessible sources, including GitHub repositories \cite{smartbugs2022curated, owasp_smartcontract} and Etherscan. From the collected pool, we select 50 representative contracts (our samples) as the basis for subsequent analysis (see \S\ref{sec:sc-collection}). Functional features of these contracts, described in \S\ref{sec:feautre-ext}, are extracted and used to construct input prompts for LLM-based agents.

To generate neutral, use-case-oriented prompts, GPT-4 is employed to mutate extracted features and produce prompt variants (see \S\ref{sec:prompt-engineering}). These prompts specify only the intended functionality of the desired contracts and deliberately avoid constraining implementation details, allowing coding agents to independently determine the contract structure.
Using these prompts, we generate smart contracts with three commercial LLM-based models: GPT-4, Gemini-2.5, and Sonnet-4 (details in \S\ref{sec:agent-code-generation}).

The generated contracts are then analyzed using Slither, a static analysis tool for Solidity~\cite{mhref09}, to evaluate \textit{correctness} and identify potential \textit{vulnerabilities}. The resulting analysis reports are aggregated and processed using custom scripts written in Python, R, and PowerShell, enabling systematic comparison and interpretation of the results.

\subsection{Smart Contract Collection}\label{sec:sc-collection}

Prior work evaluating the code-generation capabilities of LLMs has largely relied on the HumanEval dataset and its variants \cite{heibel2024mapping, liu2023chatgptcode, jenko2024blackbox}. However, HumanEval does not include Solidity, the primary language used for developing Ethereum smart contracts. To address this limitation, we curate a collection of Solidity smart contracts from three well-established sources: the SmartBugs dataset \cite{smartbugs2022curated}, the OWASP Top 10 Smart Contract Vulnerabilities \cite{owasp_smartcontract}, and the Smart Contract Weakness Classification (SWC) Registry \cite{swc2022registry}.
SmartBugs aggregates vulnerable smart contracts from Etherscan and other public sources, while the OWASP and SWC datasets contain sample codes that illustrate common vulnerability patterns in smart contract implementations. As our study focuses on scenarios in which insecure coding practices lead to exploitable vulnerabilities, we intentionally select vulnerable contracts during this stage of data collection.
Table~\ref{table:vuln-sample} summarizes the distribution of vulnerability categories represented in our sampled contracts. We note that the original datasets employ differing, and in some cases outdated, vulnerability taxonomies. For example, the SWC Registry has not been actively maintained since 2020, while SmartBugs is organized according to the Decentralized Application Security Project (DASP) taxonomy, which partially overlaps with more recent OWASP (2023) classifications.

To ensure relevance and consistency with current security practice, we align our analysis with contemporary smart contract vulnerability categorizations. As a result, the vulnerability labels presented in Table~\ref{table:vuln-sample} do not always map directly to the vulnerability categories reported in the detection and analysis phase of our study.

\begin{table}[t]
\centering
\caption{Distribution of vulnerability types in sampled smart contracts}
\label{table:vuln-sample}
\begin{tabular}{lr}
\hline
\textbf{Vulnerability Type} & \textbf{Count} \\
\hline
Access Control            & 15 \\
Denial of Service         & 6  \\
Insecure Randomness       & 3  \\
Unchecked External Calls  & 6  \\
Lack of Input Validation  & 2  \\
Logic Errors              & 3  \\
Arithmetic Issues         & 6  \\
Reentrancy                & 9  \\
Price Oracle Issues       & 2  \\
\hline
\end{tabular}
\end{table}

\subsection{Feature Extraction}\label{sec:feautre-ext}
\begin{figure}[h]
    \centering
    \includegraphics[width=1\linewidth]{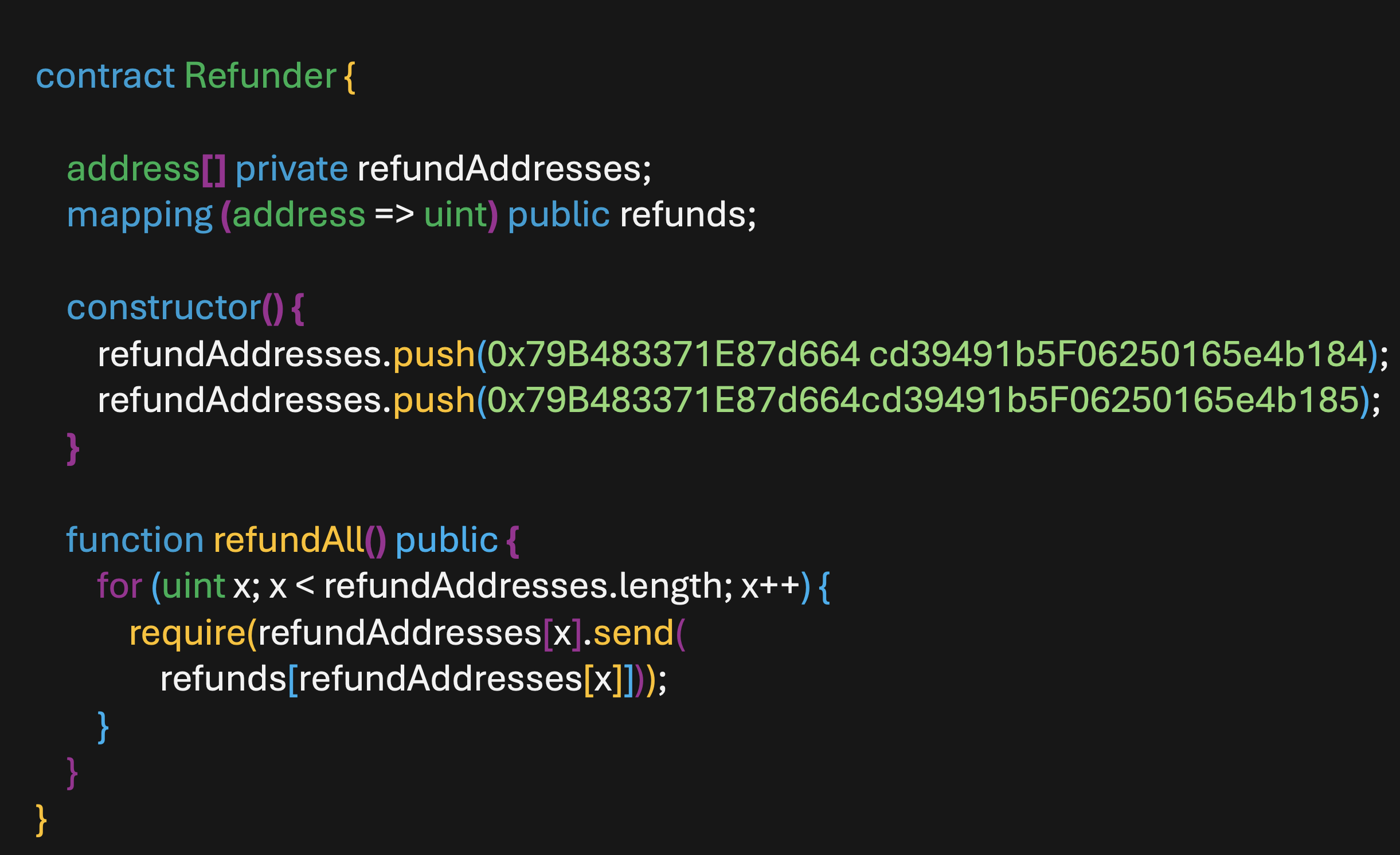}
    \caption{A sample smart contract used in Analysis}
    \label{fig:sample-xample-1}
\end{figure}

In this work, we define \emph{feature extraction} as the process of distilling the high-level functional intent of a smart contract from its source code and associated documentation. Rather than extracting low-level syntactic or structural features, our approach focuses on capturing \textit{semantic features}, such as \textit{contract purpose} and \textit{core functionalities,}, and representing them as natural language prompts. These prompts serve as a functional abstraction of the original contract and are used to elicit semantically similar implementations from LLM-based coding agents. To perform this extraction, we leverage GPT-4.1 to analyze the functions and inline documentation of each sampled contract and infer its intended functionality and application. The model produces \textit{high-level natural language descriptions} that can be directly used as prompts for LLM-based coding agents.
For example, given the sampled contract shown in Fig.~\ref{fig:sample-xample-1}, GPT-4.1 derives the prompt presented in Listing 1.~\ref{box:feature-sample}. Importantly, our objective is not to guide the coding agent toward a specific implementation strategy. Instead, to reflect realistic interactions by inexperienced developers, we deliberately exclude any instructions that prescribe how the contract should be implemented. The prompts are therefore restricted to describing the desired functionality and behavioral intent of the contract.
The final prompts used in our experiments follow this constrained formulation, as illustrated in Listing 2.~\ref{box:feature-end}. For each sampled contract, we extract two semantic features, namely, the contract’s intended purpose and its functions, which are explicitly encoded in the prompt. This abstraction preserves the contract’s high-level semantics while requiring careful interpretation by the coding agent, mirroring scenarios in which subtle logic errors may arise. The dataset consists of 52 prompts corresponding to 52 sampled smart contracts.

\begin{tcolorbox}[
    enhanced,
    colback=blue!8,
    colframe=blue!8,
    boxrule=0pt,
    arc=0pt,
    left=6pt,
    right=6pt,
    top=6pt,
    bottom=6pt,
]
\label{box:feature-sample}
\small
\textbf{Listing 1. Generated Prompt from sample Smart Contract.}
\begin{itemize}
    \item Manage a list of addresses eligible for refunds.
    \item Store refund amounts for each address in a mapping.
    \item Initialize the contract by adding two addresses to the refund list.
    \item Iterate through stored addresses and transfer the corresponding refund amounts.
    \item Revert the transaction if any refund transfer fails.
\end{itemize}
\end{tcolorbox}

\begin{tcolorbox}[
    enhanced,
    colback=blue!8,
    colframe=blue!8,
    boxrule=0pt,
    arc=0pt,
    left=6pt,
    right=6pt,
    top=6pt,
    bottom=6pt,
]
\label{box:feature-end}
\small
\textbf{Listing 2. Features Used in the Final Prompt.}
\begin{itemize}
    \item Manage a list of addresses eligible for refunds.
    \item Revert the transaction if any refund transfer fails.
 \end{itemize}
\end{tcolorbox}

\subsection{Scaling up Prompt Datasets}\label{sec:prompt-engineering}

The functional features extracted from the sampled contracts are submitted to the GPT-5-mini model via the ChatGPT interface to generate multiple prompt variants. This step is designed to increase prompt diversity while preserving the original functional intent of each contract. Prior work has demonstrated that LLM-assisted prompt mutation is effective for exploring a broader input space and revealing behavioral differences in code-generation systems \cite{yao2024fuzzllm, jenko2024blackbox, liu2023chatgptcode}.

In addition to generating paraphrased variants, we instruct the model to rephrase prompts using language representative of a junior developer. This design choice reflects realistic usage scenarios in which inexperienced programmers rely on LLM-based assistants to implement high-level ideas without precise technical specifications. By avoiding overly formal or security-aware phrasing, the resulting prompts more accurately capture the ambiguity and incompleteness typical of real-world developer requests.

This process yields a total of 34 distinct prompts, which serve as inputs to all subsequent experiments. In retrospection, we want to leverage this experiment result to find out whether contract type has any impact to the number of vulnerability in generated contract. To enable structured analysis and controlled comparison, we categorize the prompts into seven functional domains: \textit{decentralized financial services}, \textit{token management}, \textit{governance}, \textit{communication}, \textit{identity management}, \textit{trading}, and \textit{utility}. These categories capture common classes of smart contract applications and allow us to examine whether vulnerability patterns vary across functional domains.

Table~\ref{table:prompt-category} summarizes the prompt categories and provides representative descriptions for each class. The complete prompt generation procedure, including feature mutation and prompt filtering, is formalized in Algorithm~\ref{alg:promptgen}.

\begin{algorithm}
	\caption{Optimizing Prompts}
    \label{alg:promptgen}
	\begin{algorithmic}[]
        \Procedure{OptimizingPrompts}{$D_{sampled}$}
        \State \textbf{Input: } Sampled smart contracts := $D_{sampled}$
        \State \textbf{Output:} Prompts dataset := $D_{prompt}$. 
			\State Extract functions and purposes from $D_{sampled}$
            \State $D_{prompt}$ := GPT5-mini mutation output  
            \State $D_{prompt}$.\textit{length} = 34  
        \EndProcedure
	\end{algorithmic} 
\end{algorithm}

\subsection{Agent-Driven Code Generation} \label{sec:agent-code-generation}

The agent-driven code generation procedure is formalized in Algorithm~\ref{alg:generate}. Given a dataset of prompts, the algorithm submits each prompt to a selected LLM-based coding agent and requests the generation of a corresponding smart contract. During this process, the agent is required to adhere to a predefined context (Listing 3.~\ref{box:context}), which specifies the development environment and task constraints. Each generated contract is stored along with relevant metadata, including the \textit{generating model}, \textit{lines of code (LoC)}, \textit{iteration index}, and \textit{contract category}, to support subsequent analysis.

To reduce operational costs while preserving realism, we interact with three commercial LLM-based coding agents, GPT-4.1, Sonnet-4.5, and Gemini-2.5, via the Visual Studio Code GitHub Copilot plugin. While requests to GPT-4.1 are unrestricted, premium requests to Sonnet-4.5 and Gemini-2.5 are metered at a one-to-one ratio. As a result, the total cost of our experiments corresponds to a single month of a GitHub Copilot Pro Plus subscription. Our experimental design is inspired by prior black-box evaluation methodologies \cite{jenko2024blackbox}. All models are queried using identical prompts and contextual constraints to ensure fair and consistent comparison.
For each prompt, the system initializes a new project directory containing a README file populated with the prompt text, modeling the creation of a fresh development project by a programmer. We then manually interact with the coding agent through the chat interface using the predefined context, \ref{box:context}. The agent generates the complete contract, after this the process is repeated for the next prompt.

To ensure output diversity and mitigate stochastic effects, we repeat the code generation process ten times per model across the full set of 34 prompts. Each request is submitted using a one-shot interaction, ensuring that generations are isolated and unaffected by prior prompts or responses. Generated contracts are organized using a structured directory naming scheme encoding the model identity and iteration index, which facilitates automated vulnerability analysis via GitHub Actions. Additionally, LoC statistics are recorded in a CSV file for downstream quantitative analysis.

In total, this process yields a dataset of 1,033 smart contracts generated by three LLM-based coding agents.
\begin{algorithm}
	\caption{Generate Smart Contracts} 
    \label{alg:generate}
	\begin{algorithmic}[]
        \Procedure{GenerateSmartContracts}{$D_{prompt}$, $M$}
        \State \textbf{Input:} $D_{prompt}$, list of models:= $M$
        \State \textbf{Output:} contracts dataset := $D_{contract}$
        \State \textbf{for each} $model$ in $M$
        \Repeat
            \For{$prompt$ in $D_{prompt}$}
            \State Create folders with instruction in README.
            \State Provide context prompt for coding agent. 
            \State Generate smart contract using $model$
            \State \textbf{Output:} contract code, model, LoC, iteration.
            \EndFor
            \Until{10 iteration} 
        \EndProcedure
	\end{algorithmic} 
\end{algorithm}
\begin{tcolorbox}[
    enhanced,
    colback=blue!8,
    colframe=blue!8,
    boxrule=0pt,
    arc=0pt,
    left=6pt,
    right=6pt,
    top=6pt,
    bottom=6pt,
]
\label{box:context}
\small
\textbf{Listing 3. Context Prompt.\\}
Each file in the current folder contains a prompt that should be used as an instruction to generate the corresponding smart contract. The generated implementation should be appended to the existing file without removing any original content. No explanation of the code should be provided; the task should be completed by returning a confirmation upon completion.
\end{tcolorbox}

\subsection{Vulnerability analysis of the generated code.} \label{sec:vul-analysis}
To analyze vulnerabilities in the generated smart contracts, we employ \textit{Slither}, a widely used open-source static analysis framework for Solidity. Slither provides an extensive set of classifications covering over 100 classes of known smart contract weaknesses and is commonly adopted in both academic studies and industrial audits.

Algorithm~\ref{alg:detect} presents a simplified overview of Slither’s detection pipeline. Given a generated contract, Slither first compiles the source code using the native \texttt{solc} compiler to produce a JSON-formatted Abstract Syntax Tree (AST). This AST captures the syntactic structure of the contract in a representation readable for Slither’s core analysis engines. The detectors then traverse the AST to identify potential vulnerabilities and generate a structured detection report for each input contract. We provide this format of a detection entry in a Slither report of a contract. The reference field provide valuable information when extracting vulnerability data because it uniquely identify the detection module that matches the vulnerable code (Listing 4.\ref{box:report-entry-format}).

\begin{tcolorbox}[
    enhanced,
    colback=blue!8,
    colframe=blue!8,
    boxrule=0pt,
    arc=0pt,
    left=6pt,
    right=6pt,
    top=6pt,
    bottom=6pt,
]
\label{box:report-entry-format}
\small
\textbf{Listing 4. Slither Report Entry Format.}

\textit{INFO: Detectors}

\begin{itemize}
    \item \textbf{Affected functions:} List of functions in which the vulnerability is detected.
    \item \textbf{Reference:} URL linking to the corresponding Slither documentation describing the vulnerability.
\end{itemize}
\end{tcolorbox}

We process Slither’s output using custom scripts to extract vulnerability-related information, including both detected weaknesses and non-vulnerable findings. Each detected issue is subsequently mapped to a corresponding vulnerability category, where the category names are adopted directly from the Slither documentation (e.g., \textbf{pre-declaration-usage-of-local-variables}), along with an associated severity level (\textit{low}, \textit{medium}, or \textit{high}). This mapping is performed using a predefined taxonomy implemented in \texttt{RStudio}.
The final output is a structured data frame that uniquely associates each generated contract with its \textit{originating model, iteration index, detected vulnerability type}, and \textit{severity score}.

A detailed description of how these results are used to evaluate the security properties of LLM-generated contracts is provided in Section~\ref{sec:security-eval}.
\begin{algorithm}
    \caption{Detect Vulnerability from Generated Contract}
    \label{alg:detect}
    \begin{algorithmic}[]
        \Procedure{DetectVulnerability}{$D_{contract}$}
            \State \textbf{Input:} $D_{contract}$
            \State \textbf{Output:} detection report dataset := $D_{report}$
            \For{\textbf{each} $contract$ in $D_{contract}$}
                \State solc\_compiler($contract$)
                \State \Return JSON\_AST
                \State Slither\_core(JSON\_AST) 
                \State information\_recovery()
                \State slithIR\_conversion()
                \State all\_detectors
                \State \Return detection\_report
            \EndFor
        \EndProcedure
    \end{algorithmic}
\end{algorithm}

\subsection{Cost breakdown} \label{sec:cost-breakdown}
This section details the cost incurred at each stage of the experimental pipeline. The smart contract collection phase does not incur any cost, as contracts are obtained from publicly available sources such as GitHub repositories and online resources (e.g., Ethereum). Similarly, the feature extraction stage relies on 52 non-premium requests to the GPT-4.1 model, which are free of charge. The prompt generation step also incurs no cost, as requests to the GPT-5-mini model are not billed. 

Costs are primarily incurred during the agent-driven code generation phase. In total, 34 prompts are evaluated across 10 iterations for each model, resulting in 340 premium requests per model. Requests to GPT-4.1 are billed at a zero-cost rate, whereas requests to Sonnet-4.5 and Gemini-2.5 are billed at a one-to-one premium rate. Hence, the total number of paid premium requests is 680, which is fully covered by a single month of a GitHub Copilot Pro Plus subscription.

Vulnerability analysis is conducted using Slither executed via GitHub Actions with GitHub-hosted runners. Each analysis iteration takes approximately one minute to complete. Running Slither on the full dataset of 1,033 generated contracts requires roughly 30 minutes of total execution time. Given that the generated contracts collectively take only 23.5\,MB of storage, both the computation time and storage requirements fall within GitHub’s free-tier limits, which include 2,000 runner minutes and 500\,MB of storage. 

Overall, the total cost of the experimental evaluation is \$39 USD, corresponding to a one-month GitHub Copilot Pro Plus subscription, which provides access to 1,500 premium requests across supported LLM-based coding agents.\\

\section{Experimental Evaluation} \label{sec:ExprEval}

\subsubsection{Security evaluation method} \label{sec:security-eval} We provide the method we used to evaluate the security posture of the generated contract. The following statistics reflect the performance in term of secure coding of the chosen models.

\textbf{Defining vulnerable contracts:} Solidity detection report includes many $informational$ and $optimisation$ warnings. These warnings do not guarantee vulnerability or affecting the contract's functionality. Therefore, we exclude these warnings from this evaluation. In other word, a vulnerable contract is a contract with at least one \textbf{low}, \textbf{medium} or \textbf{high} detection.

\textbf{Aggregate report data:} We drafted a mapping of vulnerability name, type and severity following the Slither official document. This mapping is then joined with 

\textbf{Measuring vulnerability spawn rate:} Given the number of vulnerable contracts $C_{vuln}$ and total number of generated contracts $C_{total}$, we calculate the rate of vulnerability existing in generated code as below:
\[
\text{vuln-ratio} = \frac{\text{$C_{vuln}$}}{\text{$C_{total}$}} \times 100
\]

\textbf{Analysing the impact of vulnerable generated code}: Statistical data from the following aspects are considered: distribution of LoC and vulnerability rate, distribution of vulnerable contracts on contract type, frequency of vulnerability by type and frequency of vulnerability by severity. We expect these figures to presents accurately the weak point of code generated by three selected coding agents in particular and smart contract in general. 

\subsection{Experimental Results}
\begin{figure}
    \centering
    \includegraphics[width=1\linewidth]{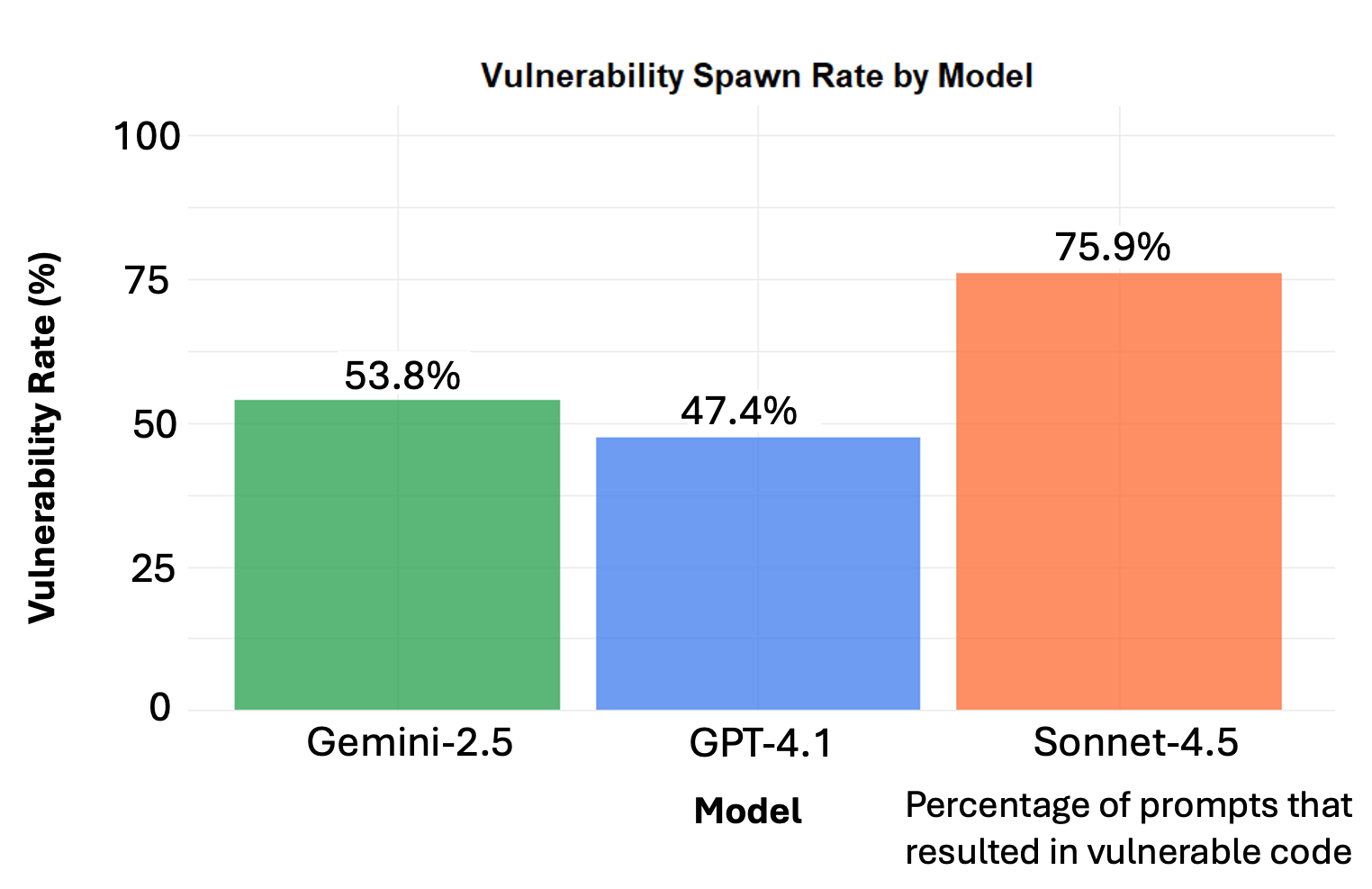}
    \caption{Vulnerability spawn rate of each chosen models.}
    \label{fig:vuln-spawn-rate}
\end{figure}

 Fig.~\ref{fig:vuln-spawn-rate} illustrates the proportion of contracts generated by the chosen models that are vulnerable. GPT-4.1 has the lowest percentage at 47.4\%, Gemini-2.5 follows closely with 53.2\%, while over 75\% of contract from Sonnet-4 is vulnerable. This means every other contract completed by a coding agent has at least one vulnerability, and for Sonnet-4, it's three out of four. The risk from unchecked use of coding agent is real and concerning.

\begin{figure}
    \centering
    \includegraphics[width=1\linewidth]{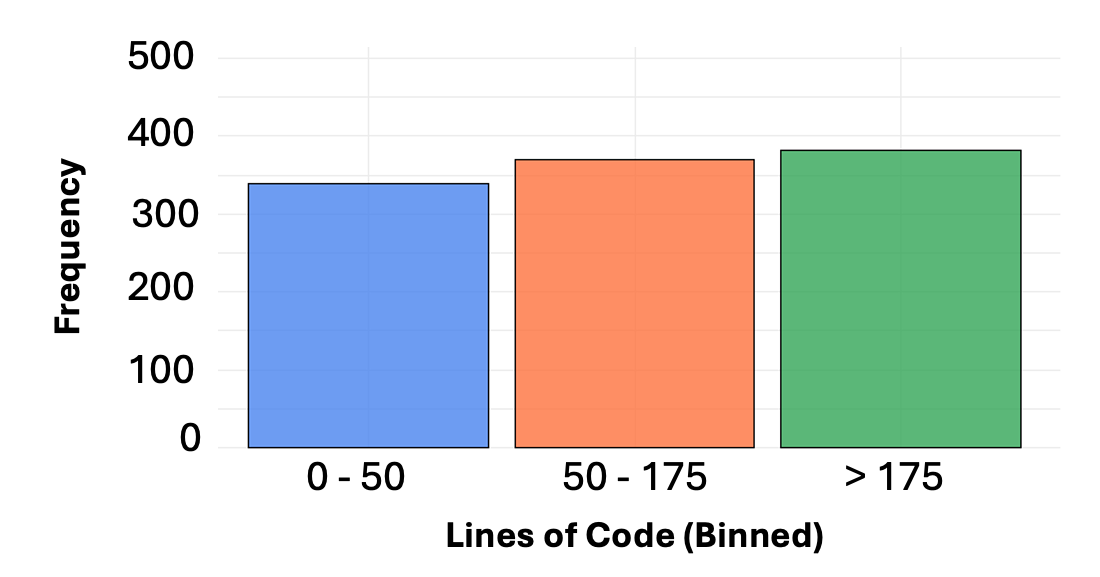}
    \caption{Line-of-Code distribution in vulnerable contracts.} 
    \label{fig:vuln-loc-distribution}
\end{figure}

The statistic of LoC data in Table.\ref{table:loc-stat} shows that at least half of the generated contracts have 75 LoC or less. Based on quartile values, Fig.~\ref{fig:vuln-loc-distribution} illustrates the positive correlation between LoC and vulnerability count using three LoC categories: 50 LoC or less, 50 to 175 LoC, and more than 175 LoC. The count of vulnerability gradually increases with contract's size. Suppose there is a chance for a line of code to be vulnerable, then the higher the LoC count, the higher the chance of a vulnerability exists in contract code. We theorized that the more code any given model generates the more vulnerability it introduces. Therefore, the count of vulnerability $freq$ was modeled as the function of contract file size, measured in Line-of-Code ($loc$) using Poisson generalised linear model with a log link. Model's output (Table.\ref{table:poisson-fit}) confirmed our theory because of the small p value. According to our model, every 100 LoC increases the count of vulnerability by 15\%.

\begin{table}[h]
\label{table:loc-stat}
\caption{Summary of LoC statistic}
\begin{tabular}{lrrrrrr}
\hline
Statistic & Min. & Q1. & Median & Mean & Q3. & Max. \\
\hline
Value     & 16.0 & 40.0    & 75.0   & 128.5 & 182.0   & 595.0 \\
\hline
\end{tabular}
\end{table}

\begin{table}[t]
\centering
\caption{Summary of lines-of-code (LoC) statistics by model}
\label{table:loc-stat-by-model}
\begin{tabular}{lrrrrr}
\hline
\textbf{Model} & \textbf{Total} & \textbf{Mean} & \textbf{SD} & \textbf{Min} & \textbf{Max} \\
\hline
Sonnet-4.5  & 69{,}987 & 222.18 & 128.48 & 28 & 595 \\
GPT-4       & 11{,}729 & 33.13  & 11.35  & 12 & 81  \\
Gemini-2.5  & 21{,}343 & 58.63  & 26.30  & 16 & 158 \\
\hline
\end{tabular}
\end{table}

\begin{table}[h]
\label{table:poisson-fit}
\centering
\caption{Output of the Poisson model.}
\begin{tabular}{lrrrr}
\hline
& Estimated & Std. Error & z-value & p \\
\hline
$loc$ & 0.0014634 & 0.0002154 & 6.793 & 1.1e-11 \\
\hline
\end{tabular}
\end{table}

\begin{figure}
    \centering
    \includegraphics[width=1\linewidth]{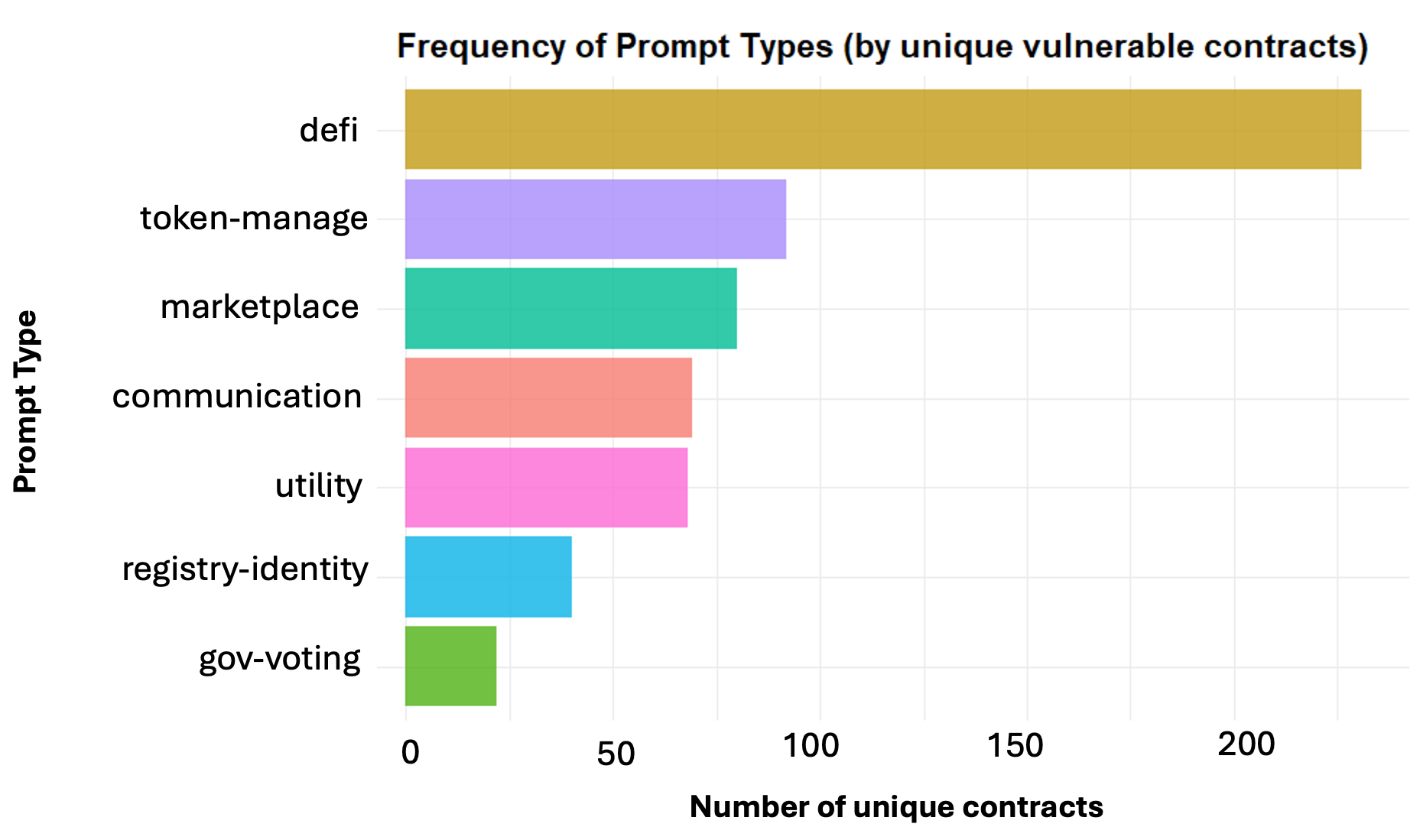}
    \caption{Vulnerability frequency in different type of contracts.}
    \label{fig:vuln-contract-by-type}
\end{figure}

Fig.~\ref{fig:vuln-contract-by-type} visualizes the frequency of vulnerability in contracts of different categories in which decentralized finance has the highest count and it exceeds category by a large amount. We found significantly less vulnerabilities from contracts of type "Government \& Voting. However, when considering the number of prompts in the pool, this result reflects the distribution of prompt type better than the trend of vulnerability existing in different contract types.

Following the classification from Slither, we count the occurrences of vulnerability by their types to find out which one are more common in LLM-generated contracts. It's worth noting that the \textit{Others} category includes many vulnerabilities that do not fit in any single category. \textit{Sonnet-4.5} tends to generate vulnerabilities in the following categories: \textit{Others}, \textit{Re-entrancy}, and \textit{Logical Error}. The figures of \textit{Sonnet-4.5} greatly overshadows the other models in those categories. In contrast, both \textit{GPT-4.5} and \textit{Gemini-2.5} generates more \textit{Lack of input validation} and \textit{Denial of Service} vulnerabilities than \textit{Sonnet-4.5}, but neither the count nor the differences between models is large. In \textit{Access Control} category, code from all three models have the same amount of vulnerabilities, with minimal gap. Interestingly, \textit{Gemini-2.5} is the only model whom code contains \textit{Unchecked External call} vulnerability. 

\begin{table}[t]
\centering
\caption{Frequency of different vulnerability types across models}
\label{table:freq-vuln-type}
\begin{tabular}{l r r r}
\hline
\textbf{Vulnerability Type} & \textbf{GPT} & \textbf{Gemini} & \textbf{Sonnet} \\
\hline
Others                     & 114 & 115 & 235 \\
Re-entrancy                & 48  & 71  & 138 \\
Lack of Input Validation   & 22  & 21  & 8   \\
Denial of Service          & 18  & 17  & 10  \\
Access Control             & 10  & 16  & 70  \\
Insecure Randomness        & 10  & 10  & 9   \\
Logic Errors               & 9   & 13  & 67  \\
Arithmetic Issues          & 1   & 13  & 27  \\
Unchecked External Calls   & 0   & 18  & 0   \\
\hline
\textbf{Total}              & \textbf{232} & \textbf{294} & \textbf{564} \\
\hline
\end{tabular}
\end{table}

We present the number of vulnerability occurrences by their severity in Table.~\ref{table:freq-vuln-severity}. This is an indicator to the impact the vulnerable contracts may cause should they reach production. The count for low severity vulnerabilities is the highest for all 3 models, and \textit{Sonnet-4.5} has the highest counts among them. For medium severity vulnerability, \textit{Sonnet-4.5} generates the most at 90 vulnerabilities while \textit{GPT-4.1} and \textit{Gemini-2.5} comes in at a much lower count, 34 and 17 respectively. In contrast, \textit{Sonnet-4} has the least count of high severity vulnerability. \textit{Gemini-2.5} created 41 high impact vulnerability, which is the largest number between the 3 models. Overall, while \textit{Sonnet-4.5} created more low and medium impact vulnerabilities than the others, its high impact vulnerability count is the least so the quality of \textit{Sonnet-4.5} contracts in term of security was partially redeemed. On the other hand, the number of vulnerability from approximately 1000 smart contracts is concerning, especially high impact ones.

\begin{table}[t]
\centering
\caption{Frequency of vulnerability severity across models}
\label{table:freq-vuln-severity}
\begin{tabular}{l r r r}
\hline
\textbf{Severity} & \textbf{GPT-4.1} & \textbf{Gemini-2.5} & \textbf{Sonnet-4.5} \\
\hline
Low    & 174 & 236 & 458 \\
Medium & 34  & 17  & 90  \\
High   & 24  & 41  & 16  \\
\hline
\end{tabular}
\end{table}

\subsection{Discussion \& Future Prospects}

\subsubsection{Implication for the use of AI agent in smart contract development.}

This experiment left a concerning remark about vulnerability spawn rate, staggering count of low impact vulnerability and significate the existence of medium and high impact vulnerability in AI-generated smart contracts. Additionally contract size was found holding meaningful statistical impact on the number of vulnerability count. This result highlights increasing needs for experienced oversight and rigorous audit. The simulation target of our experiment was inexperienced  programmers vibe-coding with coding agents. However, this potential threat could transfer to multi-agent setups where AI agents work autonomously with excessive privileges. Our experiment doesn't undermine the useful of AI agent in production but emphasizes the danger of lack of oversight and human's accountability.

\subsubsection{Limitation and Future work.}

The priority of this experiment is keeping the cost low, and reproducibility high. For that, we chose to manually conduct the data generation phase. This approach costs us more time and literally hard labor feeding the prompt to coding agent. Future study may improve on this aspect to automate the entire process, creating a larger data set with higher quality. Moreover, further improvement to vulnerability detection pipeline with Slither can be made so that it will process multiple projects concurrently. A project would have many files instead of 1 or 2 Solidity file per contract in this experiment. Lastly, We use the convenient approach and adopt Slither's vulnerability classification which left out \textit{Flash loan} vulnerability from OWASP taxonomu \cite{owasp_smartcontract}. Because Slither is a static code analyser relying on the signature of vulnerable code, it can't detect complex and market-based vulnerabilities. Future study could remedy this by revamping the whole detection process to cover the two special vulnerability categories.

\section{Conclusion} \label{conclusion}

With the rapid adoption of LLMs across the software development lifecycle, LLM-assisted code generation has become widespread in modern development workflows, including the creation of blockchain smart contracts. While this integration has significantly improved the efficiency of smart contract development, it also introduces a new class of security risks: code that is syntactically correct and functionally plausible may still harbor subtle, exploitable vulnerabilities.

In this paper, we conduct a systematic security assessment of smart contracts generated by modern LLMs, and demonstrated that syntactic correctness and functional plausibility are insufficient indicators of deployment-grade security. We evaluated contracts from several common domains, including DeFi, utilities, voting, marketplaces, and registries, generated by Gemini 2.5, GPT-4.1, and Sonnet 4.5. Our analysis reveals that high-severity vulnerabilities remain common across models and application domains. These findings indicate that LLM-generated smart contracts should be treated as untrusted code and subjected to security-first development workflows before production deployment. As future work, security-aware LLM pipelines that integrate contract generation with automated testing and formal verification could be explored to mitigate vulnerabilities before deployment.

\section*{Acknowledgement} 

This work involved limited use of large language models (e.g., ChatGPT) solely for language refinement purposes. The models were used to improve grammar and typos. The content was written entirely by the authors. No parts of the manuscript, including conceptual content, technical descriptions, data analysis, or results, were generated by any GenAI or LLMs. All scientific ideas, analyses, and conclusions are the sole responsibility of the authors.

% ----------------
% References
% ----------------
\bibliographystyle{IEEEtran} % or IEEEtran, acm, etc.
\bibliography{references}

% ----------------
% Appendices
% ----------------
%\clearpage
%\appendices
%\appendix
%\section{Additional Results}
%\input{sections/newappendix}
%

\end{document}